\documentclass[12pt,preprint]{aastex}
\usepackage{emulateapj5,apjfonts,amsmath}
\usepackage{amsmath}
\input psfig.tex

\newcommand{\kms}{$\rm {km}~\rm s^{-1}$}
\newcommand{\Msun}{M_\odot}

\begin{document}

\title{The Black-Hole Mass in M87 from Gemini/NIFS Adaptive Optics
Observations}

\author{Karl Gebhardt\altaffilmark{1},
Joshua Adams\altaffilmark{1},
Douglas Richstone\altaffilmark{2},
Tod R. Lauer\altaffilmark{3},
S. M. Faber\altaffilmark{4},
Kayhan G\"ultekin\altaffilmark{2},
Jeremy Murphy\altaffilmark{1},
Scott Tremaine\altaffilmark{5}}

\altaffiltext{1}{Department of Astronomy, University of Texas at
Austin, 1 University Station C1400, Austin, TX 78712;
gebhardt@astro.as.utexas.edu, jjadams@astro.as.utexas.edu,
murphy@astro.as.utexas.edu}

\altaffiltext{2}{Dept. of Astronomy, Dennison Bldg., Univ. of
Michigan, Ann Arbor 48109; dor@umich.edu, kayhan@umich.edu}

\altaffiltext{3}{National Optical Astronomy Observatories, P. O. Box
26732, Tucson, AZ 85726; lauer@noao.edu}

\altaffiltext{4}{UCO/Lick Observatory, UC Santa Cruz, Santa Cruz CA
95064; faber@ucolick.org}

\altaffiltext{5}{Institute for Advanced Study, Einstein Dr.,
Princeton, NJ 08540; tremaine@ias.edu}

\begin{abstract}

We present the stellar kinematics in the central 2\arcsec\ of the
luminous elliptical galaxy M87 (NGC~4486), using laser adaptive optics
to feed the Gemini telescope integral-field spectrograph, NIFS.  The
velocity dispersion rises to 480~\kms\ at 0.2\arcsec.  We combine
these data with extensive stellar kinematics out to large radii to
derive a black-hole mass equal to $(6.6\pm0.4)\times10^9~\Msun,$ using
orbit-based axisymmetric models and including only the NIFS data in
the central region. Including previously-reported ground-based data in
the central region drops the uncertainty to $0.25\times10^9~\Msun$
with no change in the best-fit mass; however, we rely on the values
derived from the NIFS-only data in the central region in order to
limit systematic differences. The best-fit model shows a significant
increase in the tangential velocity anisotropy of stars orbiting in
the central region with decreasing radius; similar to that seen in the
centers of other core galaxies.  The black-hole mass is insensitive to
the inclusion of a dark halo in the models --- the high
angular-resolution provided by the adaptive optics breaks the
degeneracy between black-hole mass and stellar mass-to-light ratio.
The present black-hole mass is in excellent agreement with the
Gebhardt \& Thomas value, implying that the dark halo must be included
when the kinematic influence of the black hole is poorly
resolved. This degeneracy implies that the black-hole masses of
luminous core galaxies, where this effect is important, may need to be
re-evaluated.  The present value exceeds the prediction of the black
hole-dispersion and black hole-luminosity relations, both of which
predict about $1\times10^9~\Msun$ for M87, by close to twice the
intrinsic scatter in the relations. The high-end of the black hole
correlations may be poorly determined at present.

\end{abstract}


\section{Introduction}

The masses of central black holes in galaxies appear to be closely
related to the luminosity (Dressler 1989; Kormendy 1993; Kormendy \&
Richstone 1995; Magorrian et al. 1998) and stellar velocity dispersion
(Ferrarese \& Merritt 2000; Gebhardt et al.\ 2000) of their host
galaxies.  These relationships, which are determined from local
samples of galaxies, provide the means to assay the cosmological mass
distribution function of massive black holes, and provide the
empirical foundation for establishing the role of black holes in
galaxy formation and evolution (e.g. Hopkins et al. 2008).

At present the black-hole galaxy-property relationships are derived
from several dozen black-hole mass determinations made over the last
few decades (see G\"ultekin et al. 2009). The relationships remain
poorly observed at both their high-mass and low-mass ends.  Lauer et
al. (2007) show, for example, that the $M_{BH}-\sigma$ and $M_{BH}-L$
relationships must be in conflict at high black-hole mass due to
curvature in the Faber \& Jackson (1976) relationship between galaxy
velocity dispersion and luminosity.  Small uncertainties in the
high-mass end of the relations can lead to uncertainties of up to two
orders of magnitude in the implied volume density of black holes with
$M_{BH}>10^9~\Msun,$ due to the high-end exponential cutoff of the
galaxy luminosity and velocity-dispersion distribution functions.
Such estimates also depend critically on knowledge of the intrinsic
scatter in the the relationships (G\"ultekin et al. 2009). Thus, there
remains a need to measure accurate black-hole masses in a sample of
the most massive galaxies.

Apart from the need to enlarge the sample of galaxies used to define
the black-hole galaxy-property relationships, it appears that we may
also need to test and potentially revise some black-hole mass
measurements already made, especially in the massive
``core galaxies.''  Recent work shows that black-hole masses are
subject to several systematic errors that have not been generally
incorporated in the models used for analyzing the data so far.  Some
of these are discussed in G\"ultekin et al. (2009) and include the
radial variations in the mass-to-light ratio due to changes in stellar
populations or the presence of a dark halo, uncertainties in the
deprojection of the surface brightness, and triaxiality, among
others. The most important of these systematic effects are:

{\it Dark halo:} Gebhardt \& Thomas (2009) show that the measured
black-hole mass for M87 increases by more than a factor of two when a
dark halo is included in the models; the reason for the change is that
the black-hole's kinematic influence is poorly resolved in the data
that they use, so that there is substantial covariance between the
black-hole mass and stellar mass-to-light ratio. In-turn the best-fit
stellar mass-to-light ratio, assumed independent of radius, is
affected by whether or not a dark halo is included in the models. It
is well understood that the mass-to-light profile for ellipticals
changes with radius and not including that trend biases the black hole
determination. An obvious, but challenging, solution to this
degeneracy is to obtain data at radii where the kinematics are
strongly dominated by the black hole rather than the stars.

{\it Incomplete orbit library:} Shen \& Gebhardt (2010) find an
increase of two in the black-hole mass for NGC~4649 when using a more
complete orbital sampling compared to models using a less coverage
(Gebhardt et al. 2003). They argue that the orbital structure near the
black hole is dominated by tangential orbits and that the older models
did not have adequate coverage of these tangential orbits (as
discussed in Thomas et al. 2004). Having too few tangential orbits
(i.e., too many radial orbits) can be compensated by having a smaller
black-hole mass.

{\it Triaxiality:} Van den Bosch \& de Zeeuw (2009) find an increase
of two in the measured black-hole mass for NGC~3379 by using triaxial
models compared to triaxial models (although they find the same
black-hole mass for M32).

All three of these systematic effects tend to increase the black-hole
mass. The increases are generally larger than the statistical
uncertainties and suggest that systematic effects still dominate.  By
observing stars close to the black hole, many model assumptions are no
longer needed. For example, if the gravitational potential is
dominated by the black hole, then the stellar contribution to the
enclosed mass is not important; hence, uncertainties in the stellar
mass-to-light ratio, which may arise from uncertainties in the
dark-halo properties, can be mitigated by probing well inside the
influence region of the black hole.

Since it is among the most luminous galaxies nearby, has the largest
black hole known (from spatially resolved kinematics) and has one of
the nearest and best-studied AGNs, M87 is a natural and important
target. An accurate black-hole mass determination for M87 helps to pin
down the sparsely sampled upper end of the black-hole mass
distribution and provides insights into formation and evolution of the
most luminous galaxies. The previous analysis of M87 from Gebhardt \&
Thomas is based on ground-based kinematic data taken in natural seeing
under moderately good conditions (FWHM=1\arcsec). In this paper, we
present kinematics based on the integral field spectrograph, NIFS, on
the Gemini Telescope, taken with adaptive optics correction. The
spatial FWHM of the kinematics is 0.1\arcsec\ on average, with the
best seeing image at 0.08\arcsec. At larger radii we incorporate new
kinematic data out to 245\arcsec\ or 2.5 effective radii that will
appear in a companion paper. The extreme improvement in the data
quality of M87 allows us to model black-hole mass with smaller
systematic uncertainty. This paper focuses on the determination of the
mass of the central black hole; the analysis of the stellar
mass-to-light ratio and the dark halo properties will be given in
Murphy et al. (2011).

Obtaining the kinematics at spatial resolution down to 0.1\arcsec, at
the same signal-to-noise obtained here, would have required about 100
orbits (90 hours) of Hubble Space Telescope, due to the faint stellar
surface brightness. This adaptive optics study using Gemini/NIFS took
about 10 hours in total, highlighting one of the great advantages for
ground-based adaptive optics.

We assume a distance to M87 of 17.9 Mpc. The value of the black-hole
mass scales linearly with assumed distance.

\section{Data}

A large amount of data exist for M87, and we do not attempt to
integrate all of it; we rely on those data that provide the highest
spatial resolution, most complete spatial coverage, and highest
signal-to-noise. Gebhardt \& Thomas (2009) combine the Hubble Space
Telescope images of Lauer et al. (1992) with the ground-based data of
Kormendy et al. (2009) at larger radii. These data determine the
surface brightness and ellipticity from radii of 0.02\arcsec\ to
2000\arcsec\ (1.7~pc to 170~kpc). Gebhardt \& Thomas deproject the
surface brightness to obtain the stellar luminosity density, and we
use their deprojected density profile in this paper. The deprojection
assumes axisymmetry; we assume an edge-on projection and the
deprojected ellipticity is generally close to zero but rises in the
central region to 0.2 and the outer region to 0.5. The stellar light
profile within 0.05\arcsec\ has large uncertainties both in the radial
shape and the ellipticity. The best-fit profile is a power law with
exponent --0.26 in radius and an increase in the ellipticity inside of
0.15\arcsec. We have run a variety of models including and excluding
this ellipticity change, increasing and decreasing the stellar power
law with a range of exponents from 0 to --0.5. We find that the
black-hole mass changes by less than the statistical
uncertainties. Thus, the exact shape of the central stellar light
profile does not appear to be important for the black-hole mass.

For the spectral data, we present new observations from the Gemini
Telescope taken with laser adaptive optics correction with the
integral field unit of NIFS. It is important to include kinematic data
out to much larger radii in order to constrain the orbital structure
and the dark halo. The main source of the stellar kinematics at larger
radii is Murphy et al. (2011). They obtain spectra with the integral
field unit on the McDonald Observatory 2.7m telescope (VIRUS-P, Hill
et al. 2009). These data extend to 245\arcsec. VIRUS-P has 4.1\arcsec\
fibers, making it unable to provide high spatial resolution. Thus, the
region between the edge of the NIFS field (radius of 1.9\arcsec) and
10\arcsec, where VIRUS-P provides adequate spatially resolved
kinematics information, requires additional coverage. We therefore
include kinematics from the SAURON integral field unit (see
http://www.strw.leidenuniv.nl/sauron). Emsellem et al. (2004) present
the SAURON data in terms of Gauss-Hermite polynomials. The dynamical
models discussed below rely on fits to the line-of-sight velocity
distribution (LOSVD). To convert the Gauss-Hermite polynomials into
LOSVDs, we generate 1000 Monte Carlo realizations of the polynomials
based on their reported uncertainties. From these realizations we
generate the average LOSVD and 68\% uncertainty at each velocity bin
used in the LOSVD.

We only use the SAURON kinematics in this region (2.5--11\arcsec) even
though they extend to 25\arcsec. We do not use SAURON data within
2.5\arcsec\ because we want to provide an independent measure of the
black-hole mass from the NIFS data alone. Including the SAURON data
within this region does not change the best-fit black-hole mass, but
does make the uncertainty smaller. We discuss these points further in
\S 6.1. At radii beyond 11\arcsec, we find a difference in the
dispersion between the SAURON values and those from the VIRUS-P data.
Murphy et al. (2011) argue that this difference is due to template
issues in the kinematic extractions, and could be related to the
limited wavelength coverage of SAURON, especially given the high alpha
enhancement of M87; see Murphy et al. 2011 for a detailed discussion
and analysis. The SAURON data used in this paper are available at the
SAURON website. 

One option to include the SAURON data at large radii would be to scale
the velocity dispersions in order to make the overlap region
consistent. We do not advocate this scaling. Primarily, the dynamical
models use the full velocity profile and not just moments, and it is
not clear whether a simple scaling of the dispersion is
adequate. Furthermore, since the offset is likely due to template
mismatch in the kinematic extraction (or continuum placement), the
scaling may not be constant with radius. Radial variations in the
scaling could be due to template mix variations with radius, velocity
dispersion changes with radius, or continuum differences with radius
due to the AGN contribution. Within 11\arcsec\ Murphy et al. find
consistency with the SAURON kinematics. Thus, we prefer to use the
SAURON data where it is consistent and exclude it where there are
differences.

We have performed several tests in which we exclude subsets of the
data---removing the SAURON data, removing some of the large radii
data, removing some of the central NIFS kinematics---and find no
effect on the best-fit black-hole mass.

\begin{figure*}[t]
\vskip 30pt \psfig{file=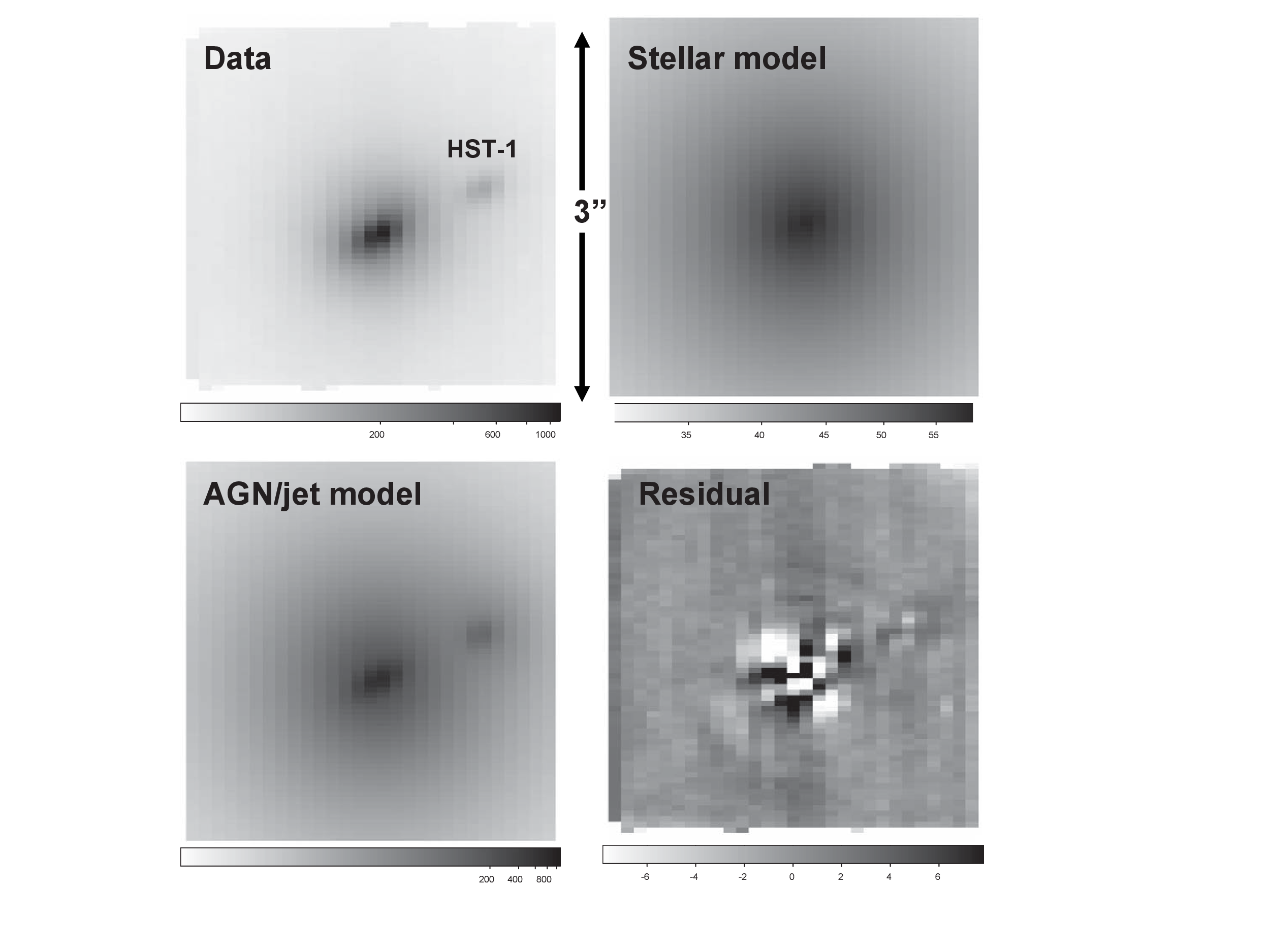,width=18cm,angle=0}
\vskip -20pt \figcaption{Stellar/AGN/clump decomposition and PSF fit
for one of the 23 datasets. In the upper left we show the original
data frame collapsed across wavelength. In the upper right, we show
the fit to the stellar distribution by enforcing a constant stellar CO
EW and spectral slope across the frame. In the lower left, we show the
fit with three point sources to the central AGN/jet, one further point
source for the HST-1 clump, and a six parameter PSF. Finally, in the
lower right we show the residuals. The final residual map is not
entirely without structure, but further point source additions do not
improve the $\chi^2$. Notice in the scales that different frames are
displayed with log and linear stretched for clarity. The scales in the
upper and lower left are the same (but arbitrary) units, and the scale
of the residuals in the bottom right are in those units. Thus, the
residuals are less than 1\%. In this image, the inner PSF is elongated
with an axis ratio of 0.6, which is can be seen in the left-side
panels.}
\label{fig_decomp}
\end{figure*}

\subsection{Gemini NIFS observations}

We observed the central region of M87 in queue mode using the
Near-infrared Integral Field Spectrograph, NIFS (McGregor et
al. 2002), on the Gemini Telescope. We used adaptive optics (AO)
corrections with the laser guide star system, ALTAIR (Boccas et
al. 2006). The AGN in M87 provided the low-order corrections for
tip-tilt and focus in a manner similar to Krajnovi{\'c} et
al. (2009). An important feature in M87's center is the nearly
point-like knot, located about 1\arcsec\ off nucleus along the
outbound jet, named HST-1 (Harris et al. 2003, Perlman et al. 2003,
Cheung et al. 2007, Madrid 2009), which allows us to monitor the
telescope's point spread function (PSF). The PSF serves as an
important input to the kinematic modeling. The data were taken over 5
nights in April and May of 2008 with 23 dithered positions of 10
minute exposure each on M87. The telluric standards HIP~59174 and
HIP~61138 were observed to monitor and correct for atmospheric
absorption. We used the K\_G5605 grating that provides wavelength
coverage from 2.00-2.43 $\mu$m, with a spectral resolution of 5290
over a field of 3$\arcsec$ $\times$ 3$\arcsec$ sampled at
0.04$\arcsec$ north-south and 0.103$\arcsec$ east-west across the
image slices.
     
We used the Gemini NIFS package of IRAF (Tody 1993) scripts (developed
mainly by T.~Beck) for the majority of the reductions. This package
provides the flattening, registration (spatially and spectrally), hot
pixel identification, and sky subtraction. This package does not yet
handle the error frames appropriately, so we also passed our original
uncertainty frames (dark current, read noise, and Poisson noise)
through the same reductions as the science data as a first
modification. Our second modification involves interpolating over the
telluric Brackett $\gamma$ line and dividing the telluric standard by
a $10^4$~K Planck function; this modification to the telluric
correction retains a proper relative spectral shape in the science
data. We also skipped the final NIFS script step where the data are
resampled to equal $x$ and $y$ steps since we found this interpolation
enhanced the residual structure from our PSF fits.

We generally took sky exposures before and after the on-target frames,
although there were some on-target frames that only had one sky
exposure. We used the sky nearest in time for each on-target
exposure. The sky subtraction usually worked well judging from
inspection of residuals under sky lines and tests with subtraction
between different pairs of sky exposures. However, it is clear that
some atmospheric emission line variability occurred between our sky
nods and caused uncertainty beyond our direct, statistical
noise. There are techniques to bundle atmospheric emission lines into
common transition families and fit a series of scalings between
science and sky frames to minimize the residuals (Oliva \& Origlia
1992, Rousselot et al. 2000, Davies 2007), but we have not employed
them here. The public CO line lists extend only to 2.27 $\mu$m and
therefore do not cover the CO bandheads through which we measure all
our kinematics. However, since we have so many sets of exposures, with
each set at a different dither position, the problematic sky regions
are mitigated. Furthermore, for the spectral extractions we mask out
wavelengths near the CO bandheads that have large variations between
sets of sky exposures above the thermal background. The final product
from the NIFS reduction package is a wavelength calibrated spectrum
for each IFU element at each of the 23 different dither positions. We
next find the relative position for each exposure, the PSF, and remove
the AGN and jet continua if present.

\subsection{Determination of the PSF, Pointings and Components}

A crucial step for dynamical modeling with AO data is to determine the
PSF and, in the case of M87, to remove non-stellar features from the
spectra. For galaxies with shallow light profiles like M87,
determination of the PSF is particularly important, since one has to
know how much light from the outer regions is in the central spatial
elements.  Fortunately, for M87 we are in the excellent situation of
having the point source (HST-1) within the field that can be used as a
measure of the PSF. However, the central AGN is so bright that it
significantly contaminates the stellar spectra within the central few
spatial elements.

There are a few techniques to estimate the PSF with AO data outlined
in Davies et al. (2004), which we considered. Davies et al. (2006)
present observations where the PSF is measured from Brackett $\gamma$
in an unresolved active galactic nucleus (AGN). Since M87 has
H$\alpha$ in the central regions, it should have some
Bracket~$\gamma$, however the redshifted line falls, to within one
pixel, on one of the brightest sky lines at 21798\AA. The residuals
from the sky line are strong enough to not detect Brackett $\gamma$
emission. The only observational handles we have on the AGN and jet
flux contributions to any particular pixel are the spatial brightness
variations and the change in spectral slope. The AGN and jet are
intrinsically much redder than the stellar populations. We wish to use
this information without making assumptions about the stellar surface
density. Thus, another goal is to study the stellar profile within the
region dominated by the AGN; using the integrated light profile does
not allow one to do this, whereas that information is contained in the
spectra.

As opposed to measuring the light model (PSF and AGN fraction) from
the reconstructed IFU data, we could use the stellar light profile as
measured from HST. There are three reasons for not forcing the light
profile from a previously measured HST image. First, the jet has knots
that are moving on short timescales, and we cannot be sure that the
AGN and jet fractional light and spatial position will be the
same. Second, we desire to use the spectral information to attempt to
measure the underlying stellar component into the center. Third, the
light profile and AGN fraction depends on the specific color. Data
with K band filters (Corbin et al. 2002) show color profiles that are
flat with radius, although the analysis cannot easily go to radii
below 1\arcsec. 

We make the assumptions of a constant stellar population into the
center, with a spatially constant spectral slope and CO equivalent
width (EW) only altered by the relative AGN/jet contamination. These
assumptions form a closed constraint on the AGN/jet components and the
PSF. Similarly to Lauer et al. (1992) with visible light HST data and
the approach of CLEAN algorithms (H{\"o}gbom 1974), we model the inner
AGN jet as a set of point sources.  Additionally, we fit PSFs to the
telluric standards as verification of our in-situ PSF models. We find
in both that the sum of an inner, AO-corrected non-circular function
and an outer, natural seeing function fit the data well without
spatially coherent residuals. The two-component PSF is common with
this type of data, but circular PSFs are commonly assumed (Neumeyer et
al. 2007).

It is important to get a robust spatial model for the AGN, jet and
stellar light since the kinematics in this spectral region are
sensitive to the equivalent width of the CO lines. Silge et al. (2005)
show that a mismatch between the equivalent widths of the velocity
templates with the data can bias the velocity dispersion either high
or low by up to 30\%. Given that the additional continuum of the AGN
will dilute the equivalent widths, it is important to use as much
information as possible to constrain the relative contribution.

Thus, we assume that i) the stellar population (and therefore color
and spectral slope) do not vary with radius near the center, ii) we
treat the AGN and jet as a set of point sources with unknown
brightnesses with a flat continuum.  We use the following operational
definition of the CO equivalent width:
\begin{equation}
\label{eq_ew}
EW_{CO}=\Delta\lambda \displaystyle\sum_{\lambda=2.29\mu m}^{2.42\mu
m} \frac{D(\lambda)-
a_{tot}\times\lambda^{\alpha_{tot}}}{a_{s}\times\lambda^{\alpha_{s}}}
\end{equation}

\noindent where $D(\lambda)$ are the counts in each pixel, $a$ is the
fitted zero-point for the continuum (from a power law fit), $\alpha$
is the fitted power law exponent for the continuum, $a_s$ and
$\alpha_s$ are the zero-point and spectral index for the stellar
light. Note that under assumption (i) this equivalent width should not
vary with position.

We begin the analysis of each science frame by considering all pixels
outside of 0.9$\arcsec$ from the center of the AGN and 0.3$\arcsec$
from the center of HST-1; in this way the fit to the stellar model use
relatively pure stellar signal.  We make a least-squares power-law fit
from 2.1--2.27 $\mu$m to each pixel and find a robust estimate for
EW$_{CO}$ and $\alpha_s$ from a biweight calculation (Beers et
al. 1990).  The estimate of any pixel's stellar continuum strength,
$a_{s}$, can then be found with a least-squares power-law minimization
for $a_{tot}$ and $\alpha_{tot}$ followed by the application of
Equation~1. The EW$_{CO}$ and $\alpha_{s}$ over all frames are
$220\pm26$\AA\ and $-3.11\pm0.43$. We make continuum and equivalent
width maps for all pixels, subtract off the non-stellar continuum, and
normalize by the stellar continuum extrapolated through the CO
bandheads. This procedure produces the reduced spectra which are later
used for extraction of the kinematics.

\subsubsection{PSF Model}

The PSF determination requires further analysis. We smooth the stellar
continuum intensities using a 0.2$\arcsec\times0.2\arcsec$ boxcar.
We assume that the PSF has the form of an anisotropic Gaussian plus a
power law (in the form of a Moffatt function), given by

\begin{align}
\label{eq_psf}
PSF(x,y) &= N(x,y) + M(x,y); \nonumber \\
N(x,y) &=
\frac{a_1}{2\pi a_2^2 a_4}
exp(-(x_c^2+\frac{y_c^2}{a_4^2})/{2a_2^2}); \nonumber \\
M(x,y) &=
(1-a_1)(a_6-1)/\pi a_5
(1+\frac{x^2+y^2}{a_5^2})^{a_6}; \nonumber \\
x_c &=x\cos(a_3)+y\sin(a_3); \nonumber \\
y_c &=y\cos(a_3)-x\sin(a_3), \nonumber \\
\end{align}

\noindent where $PSF(x,y)$ is the value of the PSF at a given position
$x$ and $y$. $N(x,y)$ is the Gaussian model for the inner PSF, with
normalization $a_1$, width $a_2$, position angle $a_3$ and axis ratio
$a_4$. $M(x,y)$ is the Moffatt model for the outer PSF, with
normalization $(1-a_1)$, width $a_5$ and exponent $a_6$. In this
model, the PSF is assumed constant over the NIFS 2\arcsec\ field; this
approximation is very accurate for a laser guide star system at a
wavelength of 2~$\mu$m.

Due to the large number of parameters needed to solve for the PSF
(shape parameters for the PSF, multiple components for the AGN/jet and
stellar profile parameters), these fitting procedures involve
minimizing a complicated function with local minima.  We resort to
simulated annealing as a global minimization tool (Press et al. 1992)
with temperature schedules that reduce by 30$\%$ over each iteration
and terminate with $10^{-4}$ fractional convergence. We first perform
a simultaneous fit to the stellar-continuum subtracted data with three
central AGN/jet point source components, a fourth point source
component for the HST-1 clump, and PSF parameters given by Equation~2.

We subsample each PSF evaluation over each pixel by 5 E-W and 3 N-S
since the individual IFU elements do not properly Nyquist sample the
PSF. This fit determines reasonable locations and strengths for all
source components, but it improperly lets the central AGN/jet drive
the PSF fit, and we know, in fact, that this feature is resolved given
the multiple components.  We therefore re-minimize the PSF terms and
the HST-1 terms with data within 1.2\arcsec\ of the preliminary HST-1
position but outside of 0.5\arcsec\ of the AGN center. The isolated
clump, HST-1, then delivers a clean PSF determination. Finally, we
reminimize all source components but hold the PSF terms fixed across
the whole map. A representative decomposition example is shown in
Figure~1 for one of the 23 datasets, where we show the raw IFU data,
the stellar model, AGN/jet model, and residuals. The bottom-right
panel shows the residuals plotted on a linear scale. There remains
some structure in the residuals, but the maximum residual is less than
1\% of the measured value. We have tried a variety of PSF models and
additional point source models, and find no improvement. Given the
tightness in the HST-1 position determination, we register all frames
off this cleanly isolated feature. From the median of all frames, we
estimate AGN and HST-1 spectral indices of $-0.67\pm0.52$ and
$-1.9\pm2.6$ respectively.

This computation also delivers the range of PSFs for the 23 datasets.
Most of the PSFs are close to the diffraction limit of 0.06\arcsec\
for the inner component, and the full range of the FWHM for the inner
component is 0.055 to 0.19. The fraction of light in the inner
component ranges from 0.14 to 0.45.. The fraction of light in the
central component is indicative of the strehl ratio; however, in
practice, the strehl ratio is hard to measure given uncertainties in
the PSF model (Gebhardt et al. 2000b). We make a two-dimensional image
of each of the 23 analytic individual PSFs. From these 23 images, we
then average to make the two dimensional array which we use for the
PSF in the dynamical modeling. 

Figure~2 plots the flux, ellipticity and position angle versus radius
for the combined PSF. The values are reported in Table~1. This PSF has
the inner Gaussian FWHM of around 0.08\arcsec\ with a fraction of the
light in the central component of $a_1=0.38$. We use the analysis of
the PSF as measured from the reconstructed IFU data, and we have also
compared this PSF to that as measured from the telluric standards. We
find a similar FWHM for the inner and outer components, but the
fraction of light in the inner component changes. For the PSF measured
from the tellurics the amount of light in the central component ranges
from 60--70\%, which is 1.5 to a factor of 2 larger than that
determined from the science frames. We argue that using the telluric
PSF is too optimistic for multiple reasons. First, the telluric star
is used as the reference star for the PSF and tip/tilt corrections
(natural guide star mode), whereas the M87 data use the laser as the
reference. Second, the M87 data use the nucleus for the tip/tilt
corrections, which is more extended than the telluric star. Third, the
M87 frames come from long exposures of 10 minutes compare to exposure
times of only a few seconds for the telluric. Thus, the M87 PSF is
expected to be more extended. In any case, we also run a full set of
dynamical models using the telluric PSF and find insignificant
differences. It is encouraging to see that both PSFs give similar
results; we attribute this robustness to the well-resolved kinematics
in the region influenced by the black hole.
 
\vskip 10pt \psfig{file=f2.ps,width=9cm,angle=0} \figcaption{The PSF
used in the dynamical modeling. The top panel is the flux versus
radius, the middle panel is the ellipticity profile, and the bottom
panel is the position angle (measured N to E). The combined PSF comes
from the sum of the 23 individual two-dimensional PSF. These values
are reported in Table~1.}
\label{fig_bin}
\vskip 10pt

\subsubsection{Central Stellar Distribution and Offset With AGN}

Bagnuolo \& Chambers (1987) and Lauer et al. (1992) both find power
law profiles with exponent $-0.26$ that stays constant into the
smallest radius measured. Subject to the constant EW and spectral
slope assumptions of our decomposition model, we investigate our
stellar profile and that as determined by Lauer et al. (1992). We find
a similar slope of $-0.2$, and find no evidence for a change in the
power law profile near the black hole.

The multiple components for the AGN and jet included in the model
provide a measure as to whether the stellar centroid is consistent
with the AGN. Batcheldor et al. (2010) report an offset of $6.8\pm0.8$
parsecs using ACS images on HST. The displacement they report is along
the jet axis, which they attribute to either a recoil event from the
black hole or a black-hole binary. The adaptive optics data presented
here have similar spatial resolution (0.075\arcsec\ for the HST data
they used versus 0.08\arcsec\ for the AO data). The very large
advantage of the AO data, however, is that the spectral information
provides a further constraint on the relative amount of AGN and
stellar contribution. The average difference between the stellar
centroid and the brightest AGN component is $3.2\pm6.0$ parsecs,
consistent with no offset. Our statistical uncertainty is 8x larger
than that of Batcheldor et al.: 0.08\arcsec\ accuracy versus
0.01\arcsec, respectively. Since the AO data should provide a better
measure of the AGN and stellar contribution, it is not understood why
the Batcheldor et al. uncertainty is 8x lower. We suspect that
important details such as the jet having multiple components (that may
move with time) and having less observational constraints (imaging
only versus imaging plus spectra) led to the result and small
uncertainty reported in Batcheldor et al. We find no evidence that the
AGN is offset from the galaxy center.

\subsection{Kinematic Fits}

To align the data with the kinematic axis at large radius from the fit
of Kormendy et al. (2009), we take a position angle of -25$\arcdeg$ (E
of N) for the major axis. Figure~3 shows the radial and angular bins
used in the modeling, following the same binning of Gebhardt \& Thomas
(2009). Figure~3 only plots the spatial region for the NIFS data (the
model goes out to 2000\arcsec), with a grayscale image from one of the
23 reconstructed IFU images. Spectra on opposite sides of the major
axis are combined. We generally have ten spectra at a given radius,
since we have 5 angular bins on each side of the minor axis. The
central two radial bins are not used due to AGN contamination
(discussed below), and the next two outer radial bins require a sum
over all angles in order to obtain adequate S/N. The total number of
spectra from the NIFS data that are used for dynamical modeling is
then 40. Table~2 provides the locations for these 40 bins. We also
include the signal-to-noise per pixel for each bin.

\vskip 10pt \psfig{file=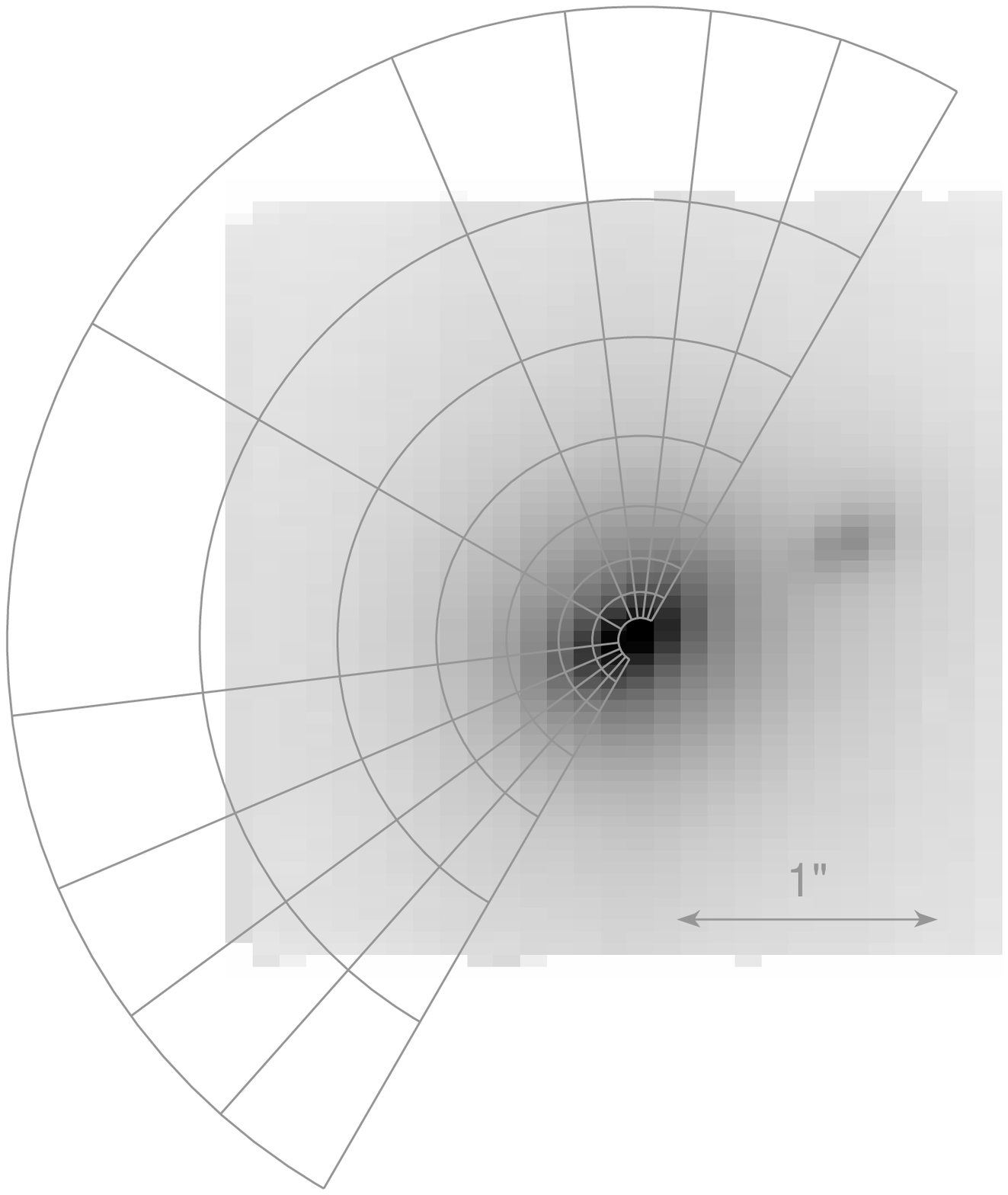,width=9cm,angle=0} \figcaption{The
binning scheme in M87 for the NIFS data only. Although this particular
frame does not have data for each bin, the dithered set fills all
bins. Data in the mirror bins around the major axis are added to the
bins shown.}
\label{fig_bin}
\vskip 10pt

Figure~4 show the spectra for three different spatial regions. The top
spectrum comes from radii $0.08<R<0.18$\arcsec, where we have summed
over all angular bins; the middle is from $0.18<R<0.3$\arcsec, and the
bottom from a radius of 0.6\arcsec. The wavelength range shown here is
the region used for the kinematic extractions. There are no
significant absorption lines on either side, although the blueward
region is used to help estimate the relative contribution of AGN and
stars (as discussed previously). The spectrum in middle panel of Fig~4
represents the innermost point used in the dynamical models. The black
line shows the data, and the dashed black lines are regions not used
in the kinematic extractions due to large variations in the night
sky. In this wavelength region, the residuals due to sky subtraction
tend to be positive, even though the subtracted sky frames use the
same exposure time as the on-target frames. We have tried different
sky subtraction levels and find little difference in the kinematic
extractions because we excise the regions with sky lines. In the other
wavelength regions, the sky residuals average to zero. The red line in
the plots is the fitted line-of-sight velocity distribution (LOSVD)
convolved with the template. The template comes from a library of 10
stars observed with GNIRS, as reported in the GNIRS template library
(Winge et al. 2009). We select stellar types from G~dwarfs to
M~supergiants. We rely on the GNIRS template library as opposed to the
NIFS template library since the GNIRS library contains a larger range
of spectral types. The template library is an important consideration
for this wavelength region (Silge et al. 2005).

\vskip 10pt \psfig{file=f4.ps,width=8.5cm,angle=0}
\figcaption[f4.ps]{Spectra at three different radii. The top is from
$0.08<R<0.18$\arcsec, the middle from $0.18<R<0.3$\arcsec, and the
bottom from $R=0.6$\arcsec. The black line is the spectrum and the
smooth red line is the best-fit template convolved with the best-fit
LOSVD. The dashed lines are those regions excluded from the fit due to
high sky contamination. The spectrum at the top, which comes from the
central region, is not used in the fit due to AGN contamination. The
velocity dispersion obtained from the fits shown in red, from top to
bottom, is 480~\kms, 480~\kms, and 445~\kms, and the S/N per pixel in
each is 30, 63, and 91.
\label{fig6}}
\vskip 10pt

The kinematic extraction simultaneously fits a non-parametric LOSVD
and template weights for the individual stars. The template
composition is allowed to vary spatially. The technique is described
in Gebhardt et al. (2000) and Pinkney et al. (2004). The LOSVD is
defined in 15 velocity bins of 260~\kms. There is a smoothing
parameter applied to the LOSVD, but given the high S/N for most of the
spectra, the smoothing has little effect on the extractions; thus,
there is only a modest correlation between adjacent velocity bins. We
use Monte Carlo simulations to determine the uncertainties in the
LOSVD. The S/N of each spectrum determines the noise to use in the
Monte Carlo simulations; from 1000 realizations of each spectrum, we
generate an average LOSVD and the 68\% uncertainty.

The dynamical modeling uses the non-parametric LOSVD
directly. However, it is sometimes convenient to express the LOSVD in
a parameterized form as Gauss-Hermite moments, to show the radial run
of the kinematics and to compare the data with the models. Table~2
shows the first four Gauss-Hermite moments for the NIFS data. Figure~5
plots the velocity dispersion versus radius, where the dispersion is
measured from a Gauss-Hermite fit to the LOSVDs. Figure~5 plots all of
the data at each radius, and there are between 1 and 10 angular bins
at each radii; thus, there are multiple points at nearly all radii in
the figure. There is no rotation seen at a significant level in the
NIFS data.

We input 107 LOSVDs in the dynamical models. These LOSVDs come from 40
spatial bins from the NIFS data, 25 from the SAURON data, and 42 from
the large radial data of Murphy et al. (2011). The data in Murphy et
al. come from the IFU VIRUS-P, where we have nearly complete angular
coverage. The S/N of those data is very high (50--100 per resolution
element). The solid line in Figure~5 plots the velocity dispersion
from the best-fit dynamical model. The model generates LOSVDs, and
their dispersions come from Gauss-Hermite fits to those LOSVDs. For
the dynamical model dispersions, we average along angles at a given
radius for clarity. In Figure~5 we plot both the NIFS and VIRUS-P
dispersions, which have different PSFs. The model is convolved to each
of the PSFs, and the plotted dispersions include the convolution.

\subsection{Spectra at R$<0.18$\arcsec}

Data within the central 0.18\arcsec\ are excluded. Within
$R<0.08$\arcsec, the number of individual NIFS spatial pixels is only
50, whereas the number of spatial pixels for bins used in the models
ranges from 250 to 3000. Given the shallow surface brightness profile
for M87 and the low number of NIFS pixels, the signal from the central
stars is low, and contamination from the AGN is high. We have tried a
variety of models for the AGN, PSF, and stellar light profile; in all
cases, we find that little information is contained in the central
spectrum. We do not further discuss this spectrum.

The spectrum coming from the radial region
0.08\arcsec$<$R$<$0.18\arcsec\ has higher S/N but still low enough
that the kinematic extraction is highly uncertain. Figure~4 plots this
spectrum in the top panel. It has lower stellar S/N compared to any
spectra we use, and is further compromised by the uncertainty in the
AGN subtraction. However, we still attempt a kinematic extraction. The
red line in Fig.~4 is the best-fit convolved template from the region
one radial bin further in radius, from $0.18<R<0.3$\arcsec. This
region has a velocity dispersion of 480~\kms.  There are wavelength
regions, for example at 2.31~$\mu$m$<\lambda<$2.35~$\mu$m, where the
model and data are offset. We do not attribute this offset to poor
LOSVD modeling but instead to poor AGN and stellar light
discrimination. We use this region as an example and we extracted
LOSVDs including and excluding various regions. The range in velocity
dispersion over all tests is 350 to 620~\kms. The uncertainties on the
dispersion for each individual extraction, coming from Monte Carlo
simulations, are much smaller than this range, indicating that we are
dominated by systematics as opposed to noise. For these reasons, we
exclude this spectrum. We note that our best-fit dynamical model
predicts a velocity dispersion of 451~\kms\ at this location (the
solid line in Fig.~5 shows the predicted dispersion from the
model). This value is in the middle of our range of dispersions using
the various extractions. For the central radial bin (R<0.08\arcsec),
the model prediction is 430~\kms.

\vskip 10pt \psfig{file=f5.ps,width=9cm,angle=0}
\figcaption[f8.ps]{Velocity dispersion versus radius for M87. The
black points are the NIFS data. The red points are the VIRUS-P data
from Murphy et al. (2011), and the blue points are from the SAURON
data. The multiple points at each radius represent the various
position angles. The solid line is the best-fit model, convolved to
the appropriate PSF. For the dynamical model, we include the predicted
dispersion within 0.18\arcsec.
\label{fig6}}
\vskip 10pt

\section{Dynamical Models}

We use the orbit-based modeling algorithm described in Gebhardt et
al. (2000, 2003), Thomas et al. (2004,2005) and Siopis et
al. (2009). These models are based on Schwarzschild's (1979) method,
and similar models are presented in Richstone \& Tremaine (1984), Rix
et al. (1997), Cretton et al. (1999), and Valluri et al. (2004).

The M87 models use the spherical geometric layout described in Murphy
et al. (2011). We use 28 radial bins and 5 angular bins for the
spatial sampling, and 15 velocity bins. The smallest spatial bin goes
from 0 to 0.05\arcsec. The velocity bins are 260~\kms\ wide. The
average number of orbits per model is 40,000. The orbital sampling
follows the design in Thomas et al. (2004, 2005), and is the same as
used in Shen \& Gebhardt (2010). The latter paper illustrates the
importance of a densely sampled orbital library: the mass obtained for
NGC~4649, a galaxy similar to M87, is a factor of 2 larger than was
found in earlier papers (e.g., Gebhardt et al. 2003). These papers did
not include enough low-eccentricity polar orbits; for those galaxies
that require significant tangential anisotropy in the central regions,
this lack of circular orbits will bias the orbital structure and hence
the black-hole mass. If a galaxy is dominated by tangential orbits in
the central regions, the projected velocity dispersion will drop (for
purely tangential orbits, the dispersion goes to zero at the galaxy
center). Thus, if the central dispersion is smaller than, for example
an isotropic distribution, this drop can be accommodated by either a
lower black-hole mass or a tangential bias with a higher black-hole
mass. In fact, the dynamical model predicts a drop in the dispersion
in the central region (solid line in Fig.~5); as discussed later this
drop is likely due to a tangential bias in the orbital distribution.

\vskip 10pt \psfig{file=f6.ps,width=9cm,angle=0}
\figcaption[f10.ps]{$\chi^2$ versus black-hole mass. Each point
represents the $\chi^2$ at that particular black-hole mass, and the
line is a smoothed curve fitted to the points. The best-fit black-hole
mass is $(6.6\pm0.4)\times10^9~\Msun$. The vertical lines represent
the 68\% range for the black hole mass. The stellar mass-to-light
ratio and the dark halo parameters have been fixed to the values
reported in Murphy et al. (2011).
\label{fig6}}
\vskip 10pt

Gebhardt \& Thomas (2009) find a strong correlation between the
black-hole mass and the circular speed of the dark halo, both of which
are anticorrelated with the stellar mass-to-light ratio. These
correlations arise because of the limited spatial resolution of their
data. This paper is based on data of higher quality, in particular
with higher spatial resolution at the center (0.1\arcsec\ compared to
1.0\arcsec) and extending to larger radii (245\arcsec\ compared to
30\arcsec\ for the stellar kinematics). Using the present data, there
are no significant correlations between the black-hole mass, stellar
mass-to-light ratio, and dark halo parameters. Therefore, we focus in
this paper on the black-hole mass, deferring a discussion of the
stellar mass-to-light ratio and dark halo parameters to Murphy et
al. (2011). Following Murphy et al., we adopt mass-to-light ratio=9.1
in $V$ (solar units) and a spherical dark halo with potential
$\Phi(r)={1\over2}V_c^2\log(r^2+R_c^2)$, $V_c=800$~\kms, $R_c=36$~kpc.

Figure~6 presents the $\chi^2$ versus black-hole mass. Each of 107
LOSVDs that we use in the dynamical models has 15 velocity bins,
sampling velocities from --1820 to 1820~\kms. Given the dispersion
profile, the outer two velocity bins at each end (i.e., 4 bins) are
zero in the models and in the data, and since the uncertainties in the
data are still significant for the large velocities, effectively they
add nothing to $\chi^2$. Thus, we have only 11 LOSVD bins that contain
signal (i.e., we could have limited the velocity range to
$\pm1400$~\kms\ with 11 bins and we would have the same $\chi^2$). The
total number of data points in the kinematic fits is therefore around
1100. There is a small correlation between the adjacent velocity bins
due to the smoothing used in the LOSVD extraction; this smoothing is
set small enough and the velocity bins are large enough (260~\kms)
that the correlation only mildly reduces the number of
degrees-of-freedom. The best-fit model has $\chi^2=848$, so the
reduced $\chi^2$ is slightly less than unity, as expected given the
correlation in the LOSVD bins. The points in Fig~6 are the $\chi^2$
values from the individual models, and the line is a smoothing
spline. We find a black-hole mass of $(6.6\pm0.4)\times10^9~\Msun$.

Figure~7 plots the ratio of the radial velocity dispersion to the
tangential dispersion for the best-fit model. The tangential
dispersion is defined as
$\sigma_t^2=0.5*(\sigma_\theta^2+\sigma_\phi^2+V_\phi^2)$, where
$\phi$ and $\theta$ are the spherical coordinates and $V_\phi$ is the
streaming motion in the $\phi$ direction. This ratio does not depend
systematically on polar angle, and so Fig.~7 plots the angular average
at a given radius. The confidence band comes from the range of models
that are within the 68\% uncertainties of the mass model, based on the
uncertainties of the four parameters: black-hole mass, stellar
mass-to-light ratio, dark halo circular velocity and dark halo core
radius. There is a sharp drop in this ratio in the center, implying a
significant amount of tangential anisotropy (or similarly a lack of
radial motion). As seen in Fig.~5, the predicted projected velocity
dispersion falls strongly inside of 0.2\arcsec\ (for orbits with no
radial dispersion, the projected dispersion in the central region
would fall to zero) due to the stronger tangential anisotropy. At
radii beyond about 30\arcsec\ the orbital structure is close to
isotropic. The tendency toward stronger tangential anisotropy in the
central region has been seen in previous analyses for other galaxies
(Gebhardt et al. 2003, 2007, Cappellari \& McDermid 2005, Shapiro et
al. 2006, Cappellari et al. 2007, Krajnovi{\'c} et
al. 2009). Theoretical models (Quinlan et al. 1995, Quinlan \&
Hernquist 1997, Milosavljevic \& Merritt 2001) predict increased
tangential anisotropy in the central regions due to a destruction of
stars on radial orbits from ejection by or accretion onto the central
black hole, leaving only those stars on tangential
orbits. Additionally, binary black holes will result in a
significantly increase tangential anisotropy (Milosavljevic \& Merritt
2001), similar to the amount seen here in M87. While it appears that
the large amount of tangential anisotropy seen here is due to a binary
black hole, a proper analysis requires a simulation tuned to the
surface brightness profile and kinematics of M87. In particular, it is
important to include a large range of initial conditions in the
simulations for the stellar orbital structure in order to use the
measured anisotropy profile to determine the evolutionary
history. Given that there are now many galaxies with well-measured
central orbital structures, this analysis would be worthwhile.

\vskip 10pt \psfig{file=f7.ps,width=9cm,angle=0}
\figcaption[f11.ps]{Radial to tangential velocity dispersion versus
radius. We average over polar angles in this plot since the variations
in $\sigma_r/\sigma_t$ between the angular bins are small. The average
ratio (solid line) and 68\% confidence band (dotted lines) come from
all models that are within the 68\% uncertainties for the four fitting
parameters (black-hole mass, stellar mass-to-light ratio, dark-halo
circular velocity, dark-halo core radius). An isotropic distribution
would have the ratio equal to unity.
\label{fig7}}
\vskip 10pt

\section{Models Without a Dark Halo}

We also ran models without a dark halo to investigate the sensitivity
of our results to assumptions about the halo. In these fits we include
kinematic data out to a radius of 100\arcsec, compared to 245\arcsec\
for the full dataset; we do not use kinematic data between 100 and
245\arcsec\ because in this region the kinematics are likely to be
dominated by the dark halo. We find that the best-fit mass decreases
to $6.4\times10^9\Msun$, only 0.5-sigma or 2\% smaller than the mass
we obtain from models with a dark halo that use all the kinematic
data. We conclude that the details of how we model the dark halo have
negligible effect on the black-hole mass. In contrast, when using data
with much lower spatial resolution (1\arcsec\ versus 0.08\arcsec\ in
the present paper), Gebhardt \& Thomas (2009) find a large change in
the black-hole mass, around a factor of 2.5, between models with and
without a dark halo. As suggested by them, the reason for this
difference is that we now have high S/N kinematic data from well
within the region influenced by the black hole, so the covariance
between the black-hole mass and stellar mass-to-light ratio is
negligible. 

\section {M87 and the BH$-\sigma$ and BH-L Relations}

The M87 black-hole mass derived here and in Gebhardt \& Thomas (2009)
is significantly larger than most of the the previous determinations
(with the notable exception of Sargent et al. 1978, which we discuss
in the next section). It is thus interesting to re-evaluate M87's
position in the correlations of black-hole mass versus velocity
dispersion (BH$-\sigma$) and versus luminosity (BH$-L$).

\subsection{The Effective Velocity Dispersion}

In G\"ultekin et al. (2009) we assign M87 a velocity dispersion of
$\sigma_e=375$~\kms, which in turn comes from the analysis of Gebhardt
et al.\ (2000).  In Figure 5, however, we see that this value is
reached only at $r<2'',$ a location that is clearly within inwardly
rising portion of the dispersion profile associated with the
black-hole's kinematic influence; this value unlikely represents the
M87 galaxy overall.

The velocity-dispersion parameter used in the BH$-\sigma$ relation is
the effective velocity dispersion, $\sigma_e$, which in Gebhardt et
al.\ (2000) is defined as $\sigma_e^2 = \int_0^{R_e} I(r)V^2(r)
dr/\int_0^{R_e} I(r) dr$, along the major axis, where $I$ is the
surface brightness and $V$ is the projected second moment of the
velocity distribution, and $R_e$ is the half light radius, which for
M87 is 100\arcsec\ as reported in Lauer et al. (2007) and Kormendy et
al. (2009). This operational definition of $\sigma_e$ appears to
provide a good correlation with black-hole mass, but there are many
different ways in which one can integrate the kinematics in order to
provide one number for the galaxy. With this definition
$\sigma_e=352$~\kms, using the kinematics and surface brightness
profile presented in this paper. The previous value of 375~\kms\
results from using the older kinematic and light profiles. It is clear
that $\sigma_e$ contains a substantial contribution from the light
inside where the black hole influences the kinematics.  If instead, we
exclude radii within this region (defined as $r_s=GM_{BH}/\sigma^2$
and equal to 2.1\arcsec\ for our models) from the integral that
determines $\sigma_e$, we find $\sigma_e=324$~\kms, about 8\%
smaller. We choose 324~\kms\ as our best estimate of $\sigma_e$, with
a range from 312~\kms\ to 352~\kms.

Churazov et al. (2010) show that there exists a radial ``sweet spot''
where the velocity dispersion at that radius is robustly related to
the circular velocity. By providing a dispersion value that is
indicative of the galaxy as a whole, this estimate may correlate well
with the black hole. Based on the kinematics from Murphy et al. (2011,
the dispersion value of the ``sweet spot'' for M87 is
$312\pm10$~\kms. Furthermore, Cappellari et al. (2007) measure a value
of 306~\kms\ by integrating the two-dimensional data within a radius
of 30\arcsec. There are a variety of ways to represent a velocity
dispersion for a galaxy, and until there is a physically-motivated
model it is not obvious which measure it optimal. Thus, in order to be
consistent with uses of $\sigma_e$ for other galaxies and the current
incarnation of the black-hole sigma correlation, one should use
$\sigma_e$ as reported above (324~\kms), but other correlations should
be studied.

We note that contamination of $\sigma_e$ by the light from stars
within black hole's kinematic influence is likely to be less important
for most other galaxies, since M87 is both close and has an unusually
massive black hole. At the same time, it may be prudent that this
issue be examined for all galaxies in the context of refining the
BH-$\sigma$ relation overall.

\subsection{Estimated Black-Hole Mass in M87}

G\"ultekin et al. (2009) present two BH-$\sigma$ relations, one for
all galaxies, and one for elliptical galaxies, alone.  Using
$\sigma_e=324^{+28}_{-12}$~\kms\ gives
$\log(M_{BH})=9.0^{+0.4}_{-0.2}$ in the first case, and
$\log(M_{BH})=9.1^{+0.4}_{-0.2}$ in the second.  Likewise, evaluating
the G\"ultekin et al. BH$-L$ relation with $M_V=-22.71$ (Lauer et al.\
2007) gives $\log(M_{BH})=9.0\pm0.2.$ Both relations thus give
$M_{BH}=1\times10^9~\Msun$, in contrast to our determination of
$M_{BH}=6.6\times10^9~\Msun$. Thus our measurement differs from the
predictions of this BH$-\sigma$ and BH$-L$ relations by 0.82
dex. However, the intrinsic scatter in these relations is estimated by
G\"ultekin et al. to be 0.44 (BH$-\sigma$, all galaxies), 0.31
(BH$-\sigma$, early-type only), and 0.38 (BH$-L$, ellipticals
only). Adding this scatter in quadrature gives estimates of
log($M_{BH})=9.0^{+0.6}_{-0.5}, 9.1^{+0.5}_{-0.4}, 9.0\pm0.4$,
respectively. Thus our measured value of log($M_{BH})=9.82\pm0.03$
differs from the predictions by 1.4--2 sigma. Given the present
uncertain state of knowledge of the high-mass ends of both the
BH$-\sigma$ and BH$-L$ relations, we do remark on the larger
significance of M87 deviation from the relations, except to say that
is does highlight the need to improve the sample of black-hole mass
determinations from the most massive galaxies.

\section{Discussion}

\subsection{M87 Specific Results}

Our best-fit black-hole mass is $(6.6\pm0.4)\times10^9\Msun$.  Sargent
et al. (1979) report a black-hole mass of $6\times10^9\Msun$ (after
scaling to our assumed distance), which is within 1-sigma of our
reported value. Their model is based on lower spatial resolution data
(about 1.5\arcsec), assumes that the velocity distribution is
isotropic, and does not include a dark halo. It is impressive that
after three decades of improvement in data quality, modeling, and
understanding, there is essentially no change in the measured
black-hole mass. Part of the reason for the robustness of the Sargent
et al. result is that the radial influence on the projected kinematics
from the black hole extends to nearly 10\arcsec\ (see Fig.~5), so the
influence of the black hole was clearly visible in their kinematic
data. They also use isotropic models, whereas we run axisymmetric
models with no restrictions on the anisotropy. To study the effect of
the assumption of isotropy, we fit isotropic models to the kinematic
data presented in this paper. The comparison between the projected
dispersions of the isotropic models and the data is poor, with an
increase in $\chi^2$ by over a factor of two. The poor fit makes it
difficult to assign a best-fit mass and the range of equally poor
fitting models have black-hole masses that range from
$6-8\times10^9~\Msun$, consistent with the models of \S3, which show
significant tangential anisotropy (Fig. 7). Thus, in M87, the
assumption of isotropy does not have a significant effect on the
measurement of the black-hole mass, although isotropic models provide
a poor fit to the data. Sargent et al. also do not include a dark
halo, which has been shown to cause the black hole to be
underestimated. Their velocity dispersions at large radii are lower
than ours (245 compared to 300~\kms), which is most likely because
their template library was incomplete and their spectra had lower
S/N. The lower dispersion causes the assumed mass-to-light ratio of
the stars to be lower, an error of the opposite sign to the error
caused by neglect of the dark halo. Thus, the impressive agreement
between our value and that of Sargent et al. (1978) appears to be due
in part to the competing effects of observational errors (dispersions
too small, which makes the stellar mass-to-light ratio too low and the
black-hole mass too large) and oversimplified models (no dark halo or
velocity anisotropy, both of which make the black-hole mass too
small). Another often-quoted black-hole mass determination from
stellar kinematics comes from Magorrian et al. (1998) who report a
value of $4.2\times10^9\Msun$ (for our distance). The likely reason
for the difference is that they do not include a dark halo and thus
overestimate the stellar mass-to-light ratio.

The black-hole mass reported here is nearly the same as that reported
in Gebhardt \& Thomas (2009), within 4\%. There is very little
kinematic data in common between the two studies. The kinematic data
in Gebhardt \& Thomas come from older long-slit data at spatial
resolution of 1.0\arcsec\ (van der Marel 1994), while in this paper we
use two-dimensional coverage at spatial resolution of 0.08\arcsec. We
further use ground-based data from Murphy et al. (2011) that have
excellent S/N and radial extent. There is some data from SAURON
(Emsellem et al. 2004) in common between the two studies, but this
provides only 10\% of the LOSVDs used in the models. Thus, the
dynamical models from the two studies use nearly independent kinematic
datasets, and give approximately the same answer.

The uncertainties on the black-hole mass from these two studies are
similar even though the data presented here are superior in many ways;
the previous uncertainty is $0.5\times10^9$ whereas the uncertainty
with the NIFS data is $0.4\times10^9$. In order to keep the black-hole
mass measures independent, the models presented in this paper do not
include the SAURON data inside of 2.5\arcsec. The similarity in the
black-hole mass uncertainty is then due primarily to the fact that the
two sets of data have similar accuracy on the kinematics in the
central 2.5\arcsec. Combining all NIFS data, the accuracy on the
velocity dispersion is 0.2\% (1~\kms). Combining all SAURON data
within 2.5\arcsec\ provides the same accuracy. Thus, as long as one
has a reliable PSF and no systematic differences in the kinematic
extractions, then it is expected that the uncertainty on the
black-hole mass is similar using either dataset. We have run a subset
of models including both the NIFS data and all SAURON data; in this
case, the uncertainty on the black-hole mass decreases to
$0.25\times10^9$ (with no change in the best-fit mass). We report and
utilize the result using only the NIFS data within 2.5\arcsec\ in
order to 1) provide as independent result as realistically possible
and 2) control potential systematic differences in the kinematic
extractions. Murphy et al. (2011) find a difference in the velocity
dispersion of the SAURON data at large radii compared to their
measurements, which they attribute to template issues. While we do not
find an offset in the dispersion values in the central region, we
desire to maintain the independence. The major difference, however, is
that there is no degeneracy with the stellar mass-to-light ratio using
the NIFS data, whereas the degeneracy is very strong otherwise. Thus,
the systematic uncertainty from the mass-to-light ratio profile is
effectively removed with the adaptive optics data, making the result
on the black-hole mass and orbital structure much more robust.

For M87, the AO data has removed the systematics due to the
mass-to-light ratio profile but the systematics due to the extraction
of the kinematics remain important. These systematics include
continuum placement, template mismatch, and removal of the AGN
contribution. The first two are general and the latter is specific to
M87. Getting any of these controlled to better than 1\% of the
velocity dispersion will be very difficult.

Corrected to our distance, the black-hole masses reported from gas
kinematics are ($2.9\pm0.8)\times10^9~\Msun$ in Harms et al. (1994)
and ($3.8\pm1.1)\times10^9~\Msun$ in Macchetto et al. (1997). As
discussed in Gebhardt \& Thomas (2009) the mass reported here is in
conflict with these by about 2-sigma. Possible reasons for the
differences are discussed in Gebhardt \& Thomas, with the most likely
reason being uncertainty in the inclination of the gas disk. Macchetto
et al. assume a value of 51 degrees based on the gas kinematics. Harms
et al. assume a value of 42 degrees based on the imaging of the gas
emission. The reported difference provides a measure of the systematic
uncertainty in the inclination (i.e., whether the gas kinematics or
the gas distribution are more affected by non-gravitational
forces). Applying this 9 degree difference in the inclination changes
the Macchetto et al. black-hole mass from $(3.8\pm1.1$) to
$(5.4\pm1.3)\times10^9~\Msun$, which would lead to an insignificant
difference of 0.6-sigma between our result and theirs. Of course, the
analysis is more complicated than this simple application since one
would need to re-model the gas kinematics with a different
inclination.  A proper treatment would be to include the gas
kinematics with the stellar dynamical models. Our focus in this paper
is on the stellar kinematics, and we do not attempt to merge the gas
kinematic analysis.

\subsection{General Implications for Black-Hole Mass Measurements}

While the kinematics obtained from the adaptive optics study produce
effectively the same black-hole mass and its uncertainty from
kinematics taken in native image quality, the robustness of the
measures is greatly strengthened. For example, the black-hole mass is
not dependent on the assumption of constant mass-to-light
ratio. Trying to generalize this result to other galaxies with
black-hole mass determinations is difficult since the measure of the
black-hole mass depends on many aspects. There are two observational
extremes that we highlight as examples. The first is having a measure
of black-hole mass that comes from observations that resolve well the
kinematic influence of the black hole. In the most extreme case, high
S/N spectra could potentially see the high velocity wings in the LOSVD
due to the black hole (as discussed in van der Marel 1994). The second
example would be to allow poorer resolution of the black hole but
provide a very accurate measure of the mass-to-light profile. In this
paper, we rely on the first strategy; Gebhardt \& Thomas rely on the
second. That the two strategies give consistent results, at least for
M87, suggests that both may be reliable.

Other studies have reported robust measures of the black-hole mass
from ground-based studies that only poorly resolve the black hole's
kinematic influence. Shapiro et al. (2006) measure a black-hole for
NGC~3379 from SAURON data that is consistent with that measured from
{\it HST} data using both stars (Gebhardt et al. 2000c) and gas
kinematics. Kormendy (2004) summarizes the history of black-hole mass
measures for many galaxies and finds that, in general, the differences
are within the reported uncertainties. If one has sufficient
signal-to-noise and two-dimensional coverage (e.g., SAURON or
VIRUS-P), then it should be possible to measure a black-hole mass
robustly. Thus, it is not necessarily required to resolve the region
influenced by the black hole.

Being able to use data that does not well resolve the black hole's
influence on the kinematics allows us to study black holes that are
either distant or low mass. Both of these regimes are important for
understanding the physical nature of the black hole correlations with
the host galaxy. For example, McConnell et al. (2011) measure a black
hole mass in NGC~6086, which is 133 Mpc distant. The kinematic
influence of the black hole is barely resolved, and the degeneracy
between the black hole mass and M/L profile is strong. However, as
demonstrated for M87, as long as one properly characterizes the mass
profile at large radii, then high signal-to-noise data can measure the
black hole mass accurately.

It is possible that systematic uncertainties bias the current crop of
black-hole correlations. One obvious consequence could be that without
accounting for the effect of systematic uncertainties, the measured
intrinsic scatter would increase. G\"ultekin et al. (2009) measure
scatter of 0.44 dex for the full sample of galaxies with measured
black-hole masses and 0.31 dex for ellipticals. Once systematic
effects are understood and included, the intrinsic scatter may
decrease. Other consequences include increasing the mass density of
black holes, if black-hole masses are all underestimated, and changing
the slope or curvature of any correlation. Schulze \& Gebhardt (2011)
re-analyse the set of 12 galaxies from Gebhardt et al. (2003)
including a dark halo. They find an increase of 50\% in the black-hole
mass, due primarily to improved dynamical modeling (more complete
orbit sampling) and partly to including a dark halo. The increase
correlates with black-hole mass. It is important to re-evaluate all
black-hole mass estimates.

The key to understanding all of these effects comes from high spatial
resolution data. Data from Hubble Space Telescope (mainly from STIS)
is generally regarded as providing the most significant results for
black-hole mass studies. The small and stable PSF is a central aspect
for the robustness of the data from HST. Future uses of STIS will play
an important role for quantifying black-hole masses. The main obstacle
for HST though is that it is a relatively small mirror and requires
substantial observing time. For example, in order to measure the
black-hole mass in M87 at the same accuracy presented here would
require nearly 100 orbits. While this amount of time could be
justified for a small number of objects, going to a much larger sample
using HST is difficult. Fortunately, adaptive optics observations are
in a mature stage where they can provide much larger samples.

\acknowledgements

We are grateful to the excellent staff at the Gemini Telescope, both
for providing the observations and significant support in the data
analysis. The NIFS spectrograph with the laser adaptive optics system,
ALTAIR, is a powerful resource. We thank the referee for excellent and
thoughtful comments which improved the manuscript. KG greatly
acknowledges Ralf Bender, the Max-Planck-Institut fuer
Extraterrestrische Physik and Carnegie Institute of Washington for
their excellent support and hospitality. The project would not have
been possible without the facilities at the Texas Advanced Computing
Center at The University of Texas at Austin, which has allowed access
to over 5000 node computers where we ran all of the models. KG
acknowledges support from NSF-0908639. The observations were obtained
at the Gemini Observatory, which is operated by the Association of
Universities for Research in Astronomy, Inc., under a cooperative
agreement with the NSF on behalf of the Gemini partnership: the
National Science Foundation (United States), the Science and
Technology Facilities Council (United Kingdom), the National Research
Council (Canada), CONICYT (Chile), the Australian Research Council
(Australia), Minist\'{e}rio da Ci\^{e}ncia e Tecnologia (Brazil) and
Ministerio de Ciencia, Tecnolog\'{i}a e Innovaci\'{o}n Productiva
(Argentina), under program GN-2008A-Q-12. The paper uses data obtained
at the William Herschel Telescope, operated by the Isaac Newton Group
in the Spanish Observatorio del Roque de los Muchachos of the
Instituto de Astrof\'{i}sica de Canarias


\end{document}